\documentstyle[preprint,aps,prb,eqsecnum]{revtex}

\begin{document}
\draft
\title{Interaction of a Magnetic Impurity with Strongly
Correlated Conduction Electrons}
\author{Tom Schork and Peter Fulde}
\address{Max-Planck-Institut f\"ur Physik komplexer Systeme,
Bayreuther Str.\ Hs.\ 16, D-01187 Dresden, Germany}
\date{\today}
\maketitle

\begin{abstract}
We consider a magnetic impurity which interacts by hybridization
with a system of strongly correlated conduction electrons. The
latter are described by a Hubbard Hamiltonian. By means of a
canconical transformation the charge degrees of freedom of the
magnetic impurity are eliminated. The resulting effective
Hamiltonian $H_{\rm eff}$ is investigated and various limiting
cases are considered. If the Hubbard interaction $U$ between
the conduction electrons is neglected $H_{\rm eff}$ reduces to a
form obtained by the Schrieffer-Wolff transformation, which is
essentially the Kondo Hamiltonian.  If $U$ is large and the
correlations are strong $H_{\rm eff}$ is changed.  One
modification concerns the coefficient of the dominant exchange
coupling of the magnetic impurity with the nearest lattice
site. When the system is hole doped, there is also an
antiferromagnetic coupling to the nearest neighbors of that site
involving additionally a hole. Furthermore, it is found that the
magnetic impurity attracts a hole. In the case of electron
doping, double occupancies are repelled by the impurity. In
contrast to the hole-doped case, we find no magnetic coupling
which additionally involves a doubly occupied site.
\end{abstract}


\narrowtext

\section{Introduction}
\label{sec:intro}

Recently, heavy-fermion behavior has been observed in the
electron-doped cuprate Nd$_{2-x}$Ce$_x$CuO$_4$
\hbox{($0.1\lesssim x\lesssim 0.2$)}.\cite{Czjzek} Below 0.3~K a
linear specific heat \hbox{$C_v = \gamma T$} is observed with a
large Sommerfeld coefficient \hbox{$\gamma \simeq 4{\rm
J}/(\mbox{mole Nd}\cdot {\rm K^2})$}. In the same temperature
regime, the spin susceptibility is found to be independent of
the temperature and the Sommerfeld-Wilson ratio is of order
unity. These are characteristic features of heavy-fermion
excitations.\cite{FuldeKZ} However, the characteristic low
energy scale of the order of 1~K which is associated with this
behavior is not based on a Kondo lattice effect, as it is the
case in other heavy-fermion systems. Rather, it is based on a
Zeeman effect, a consequence of the strong electron correlations
in the conducting $\rm CuO_2$ planes.\cite{FuldeZZ} Indeed,
undoped $\rm Nd_2CuO_4$ is an antiferromagnetic charge-transfer
insulator instead of a metal,\cite{Skanthakumar,Oseroff} despite
of one hole per unit cell. The Nd ions are therefore coupling to
a system of strongly correlated
electrons\cite{Boothroyd,CzjzekII} rather than to weakly or
uncorrelated ones.

As a first step towards an understanding of the consequences of
the strong correlations, the effect of a single impurity should
be investigated. Several authors studied the influence of
non-magnetic impurities in systems with strongly correlated
electron in order to explain the magnetic properties of the
undoped or weakly doped host materials of the high-temperature
superconductors (e.g., $\rm La_2CuO_4$). The considered
impurities stem from substituting $\rm Sr$ for $\rm
La$,\cite{SzczepanskiRZ,RubinS,RabeB} or from substitutions in
the $\rm CuO_2$ planes, i.e., $\rm Zn$ for $\rm
Cu$\cite{BulutHSL,AcquaronP,PoilblancHS} and $\rm S$ for $\rm
O$.\cite{ShermanR} Nagaosa {\em et al.}\cite{NagaosaHI} treated
magnetic impurities in an undoped two-dimensional Heisenberg
antiferromagnet by adding external spins with a different
exchange coupling constant. Nagao {\em et al.}\cite{NagaoMY}
discussed the influence of both, non-magnetic and magnetic
impurity scattering on the spin density wave state and the
superconducting state within a Hubbard model for weakly
correlated conduction electrons.

In this paper, we study the coupling of a single magnetic
impurity to strongly correlated electrons moving on a
lattice. The latter are described by a Hubbard Hamiltonian and
the total Hamiltonian must, therefore, go beyond that of the
single-site Anderson impurity model.\cite{Anderson} By means of
a canonical transformation the charge degrees of freedom of the
impurity site are eliminated like by the Schrieffer-Wolff
transformation\cite{SchriefferWolff} in the case of uncorrelated
electrons. Due to the strong correlations of the conduction
electrons, new terms appear in the resulting effective
Hamiltonian. They are analyzed and interpreted. Special
attention is paid to the case of a nearly half-filled conduction
band with strong correlations, a situation prevailing, e.g., in
Nd$_{2-x}$Ce$_x$CuO$_4$. We believe that the new terms resulting
from the correlations of the conduction electrons are of
relevance for a number of different systems. If the correlations
of the conduction electrons are neglected the effective
Hamiltonian reduces again to that of Schrieffer and
Wolff.\cite{SchriefferWolff}

In the next section the starting Hamiltonian is defined and the
canonical transformation is outlined. The resulting effective
Hamiltonian for uncorrelated electrons is discussed in
Sec.~\ref{sec:smallU}. Particularly interesting is the limit of
strong correlations. In Sec.~\ref{sec:largeU}, we discuss the
system at half filling and both, the hole and the electron doped
case. Finally, Sec.~\ref{sec:summary} contains a summary of the
results and the conclusions.

\section{The Hamiltonian and its transformation}
\label{sec:Ham}

We describe the strongly correlated conduction electrons in the
substrate by a Hubbard Hamiltonian on a hypercubic lattice with
unit vectors $x$
\begin{equation}
H_{c} = - t\sum_{j x,\sigma}
c^\dagger_{j\sigma} c^{\phantom{\dagger}}_{j+x\sigma}
+ U \sum_j n_{j\uparrow} n_{j\downarrow}
= H_t + H_U .
\end{equation}
The operators $c^\dagger_{j\sigma}$
($c^{\phantom{\dagger}}_{j\sigma}$) create (destroy) an electron
with spin $\sigma$ on site $j$ and $n_{j\sigma} =
c^\dagger_{j\sigma} c^{\phantom{\dagger}}_{j\sigma}$. The
magnetic impurity is assumed to contain one orbital (e.g.,
4$f$), which is either empty or singly occupied. Double
occupancies are excluded because of the strong repulsion of
electrons in that orbital. The orbital energy is therefore given
by
\begin{equation}
H_{f} = \epsilon_f \sum_{\sigma}
\hat f^\dagger_{\sigma} \hat f^{\phantom{\dagger}}_{\sigma}
\end{equation}
with $\hat f^\dagger_{\sigma} = f^\dagger_{\sigma} (1-n_f)$,
$n_f = \sum_{\sigma} f^\dagger_{\sigma}
f^{\phantom{\dagger}}_{\sigma}$, and \hbox{$\epsilon_f<0$}. The
operators $f^\dagger_\sigma$ and $f^{\phantom{\dagger}}_\sigma$
obey the usual anticommutation relations. For later convenience,
we define \hbox{$r=-\epsilon_f/U (>0)$}. Finally, we assume that
the interaction between the $f$ orbital and the conduction
electrons is local and described by a hybridization contribution
like in the Anderson model~\cite{Anderson}
\begin{equation}
H_{c-f} = V \sum_{\sigma} \left(
c^\dagger_{0\sigma} \hat f^{\phantom{\dagger}}_\sigma +
\hat f^\dagger_\sigma c^{\phantom{\dagger}}_{0\sigma} \right) .
\end{equation}
Instead of coupling the $f$ orbital with the site 0 of the
Hubbard lattice we could have replaced one site of the lattice
by the $f$ orbital. The present choice is motivated by the
situation prevailing in $\rm Nd_2CuO_4$.

The total Hamiltonian reads
\begin{equation}
\begin{array}{r c l}
H & = & H_c + H_f + H_{c-f}\\
\rule{0mm}{6mm}
& = & \displaystyle - t\sum_{j x,\sigma}
c^\dagger_{j\sigma} c^{\phantom{\dagger}}_{j+x\sigma}
+ U \sum_j n_{j\uparrow} n_{j\downarrow}
+ \epsilon_f \sum_{\sigma}
\hat f^\dagger_{\sigma} \hat f^{\phantom{\dagger}}_{\sigma}
\\
& & \displaystyle + V \sum_{\sigma} \left(
c^\dagger_{0\sigma} \hat f^{\phantom{\dagger}}_\sigma +
\hat f^\dagger_\sigma c^{\phantom{\dagger}}_{0\sigma} \right) .
\end{array}
\label{ham}
\end{equation}

{}From this Hamiltonian we want to derive an effective one,
$H_{\rm eff}$, which acts on the space with a singly occupied
$f$ orbital, only. Configurations with an empty $f$ orbital are
eliminated by first performing a canonical transformation,
thereby eliminating the mixing of configurations with different
$f$ orbital occupancies and projecting afterwards onto the space
with $n_f=1$. The canonical transformation is written as
\begin{equation}
H_{\rm can} = e^S H e^{-S} ,
\label{hcan}
\end{equation}
where $S$ is determined by the requirement that $H_{c-f}$
disappears to lowest order in $V$. This leads to
\begin{equation}
[H_c+H_f, S]_- = (L_c+L_f) S = H_{c-f} .
\label{kans0}
\end{equation}
Here we introduced the Liouville operators $L_c$ and $L_f$ which
act on an operator $A$ according to $L_c A = [H_c,A]_-$ and $L_f
A = [H_f,A]_-$, respectively. Equation~(\ref{kans0}) has the
formal solution
\begin{equation}
S = \frac{1}{L_c+L_f} H_{c-f} ,
\label{kans1}
\end{equation}
and up to terms of second order in $V$ the Hamiltonian $H_{\rm
can}$ reads
\begin{equation}
H_{\rm can} = H_c + H_f + \frac{1}{2} [S, H_{c-f}]_- .
\end{equation}
Next, we project by means of a projector $P_f$ onto the space
with $n_f=1$.  The effective Hamiltonian is then given by
\begin{equation}
\begin{array}{r c l}
H_{\rm eff} & = &  P_f H_{\rm can}P_f
\\
& = &  \epsilon_f + H_c
\displaystyle + \frac{V^2}{2} \sum_{\sigma,\sigma'}
(\hat f^\dagger_{\sigma'} \hat f^{\phantom{\dagger}}_{\sigma}
c^{\phantom{\dagger}}_{0\sigma'} \frac{1}{\epsilon_f - L_c}
c^\dagger_{0\sigma} + {\rm hc}).
\end{array}
\label{heff1}
\end{equation}

To proceed further we use the expansion
\begin{equation}
\begin{array}{r c l}
\displaystyle\frac{1}{\epsilon_f - L_c} c^\dagger_{0\sigma}
& = &
\displaystyle\frac{1}{\epsilon_f - L_U - L_t}
c^\dagger_{0\sigma}
\\
\rule{0mm}{9mm} & = & \displaystyle
\sum_{\nu=0}^{\infty} \left( \frac{1}{\epsilon_f - L_U}
L_t\right)^\nu  \frac{1}{\epsilon_f - L_U} c^\dagger_{0\sigma},
\end{array}
\label{expand}
\end{equation}
which is one in powers of $t/\epsilon_f$, and terminate it after
\hbox{$\nu=2$}. The Liouvilleans $L_t$ and $L_U$ in
Eq.~(\ref{expand}) are defined by $L_t A = [H_t,A]_-$ and $L_U A
= [H_U,A]_-$, respectively. These terms are easily evaluated if
one expresses $c^\dagger_{i\sigma}$ in terms of eigenoperators
of $L_U$, i.e.,
\begin{equation}
c^\dagger_{i\sigma} =
c^\dagger_{i\sigma} (1-n_{i\bar\sigma})
+ c^\dagger_{i\sigma} n_{i\bar\sigma}
= \hat c^\dagger_{i\sigma} + \bar c^\dagger_{i\sigma} ,
\end{equation}
where $\bar \sigma = -\sigma$. The last two operators have the
eigenvalues $0$ and $1$, respectively. The resulting terms of
Eq.~(\ref{expand}) up to second order are listed in
Appendix~\ref{sec:general}, where next-nearest neighbor
contributions which arise in second order ($\nu=2$) have been
neglected.

Inserting these terms into Eq.~(\ref{heff1}) one obtains $H_{\rm
eff}$ in the form
\begin{equation}
H_{\rm eff}= \epsilon_f + H_c + H^{(0)} + H^{(1)} + H^{(2)} .
\label{heff2}
\end{equation}
The Hamiltonians $H^{(\nu)}$ are of order $t^\nu$. The
general expressions, which hold for arbitrary values of
$U/\epsilon_f$, are lengthy and are, therefore, moved to
Appendix~\ref{sec:general}. In the following we discuss and list
only special limits of them.

\section{Limit of uncorrelated conduction electrons}
\label{sec:smallU}

For $U = 0$ ($r=\infty$), we obtain from Eq.~(\ref{heff2}) the
result of the conventional Schrieffer-Wolff
transformation~\cite{SchriefferWolff}, i.e.,
\begin{equation}
\begin{array}{r c l}
H_{\rm eff}(U=0) & = & \displaystyle\epsilon_f
+ \frac{\gamma' V^2}{2\epsilon_f} \left( 2 - n_0 \right)
\\
&& \displaystyle + \sum_{jx,\sigma}^{\phantom{0}} T^{(0)}_j
\left( c^\dagger_{j\sigma} c^{\phantom{\dagger}}_{j+x\sigma} +
c^\dagger_{j+x\sigma} c^{\phantom{\dagger}}_{j\sigma} \right)
\phantom{\frac{V^2}{\epsilon_f}}
\\
&& \displaystyle
- \frac{2\gamma' V^2}{\epsilon_f} {\bf S}_f {\bf s}_0
+ \frac{tV^2}{\epsilon_f^2}
{\bf S}_f \sum_{i(0)}^{\phantom{0}}
({\bf s}_{i0} + {\bf s}_{0i}).
\end{array}
\label{Uzero}
\end{equation}
Here, the summation $i(0)$ is over the $Z$ nearest neighbor
sites of site $0$, ${\bf S}_f$ is the $f$ electron spin, ${\bf
s}_0$ is that of site $0$ and
\begin{equation}
{\bf s}_{i0} = \displaystyle\frac{1}{2} \sum_{\alpha\beta}
c^\dagger_{i \alpha}
{\bbox{\sigma}}^{\phantom{\dagger}}_{\alpha\beta}
c^{\phantom{\dagger}}_{0 \beta} .
\label{spin}
\end{equation}
${\bbox{\sigma}}$ denote the Pauli matrices.

The last two terms in Eq.~(\ref{Uzero}) describe the Kondo
coupling between the $f$ spin and the spin of the conduction
electrons.\cite{Kondo} Applying $L_t$ in Eq.~(\ref{expand})
results in a delocalization of the interaction to first order in
$t$. To second order, we obtain a renormalization factor
$\gamma'= 1+ Zt^2/\epsilon_f^2$ for ${\bf S}_f {\bf
s}_0$. Further delocalization of the interaction is neglected in
our approximation, since next-nearest neighbor contributions
were not taken into account in the truncated form of
Eq.~(\ref{expand}), see Appendix~\ref{sec:general}. Due to the
impurity the electron hopping is reduced between site $0$ and
its nearest neighbors
\begin{equation}
T^{(0)}_j = -\displaystyle\frac{t}{2}
\left( 1-\frac{V^2}{2\epsilon_f^2} \delta_{j0} \right) .
\end{equation}
Finally, there is a one-particle potential, which reduces the
probability of having an electron at site 0, as this would
diminish the gain in kinetic energy of the $f$ electron by a
virtual hop.

\section{Limit of strong correlations}
\label{sec:largeU}

Of particular interest is the limit of strong correlations ($U
\gg t$), which we treat in the following. In contrast to the
ordinary Hubbard model (without impurity), the two cases of
doping the half-filled system with holes or with electrons are
no longer symmetric since we take into account an empty and a
singly occupied $f$ orbital only. In the following, we perform a
degenerate perturbation theory for both cases and comment on the
half-filled case.

\subsection{Hole doping}
\label{subsec:hole}

For less than half filling (hole doping), doubly occupied sites
are excluded in the limit of strong correlations (large
$U$). Let $P_c$ denote the projector onto the configuration
space without doubly occupied sites. Furthermore, set $Q_c =
1-P_c$. The effective Hamiltonian, when reduced to the space of
no doubly occupied lattice sites becomes~\cite{Kato}
\begin{equation}
\tilde H^{(\rm h)} = P_c H_{\rm eff} P_c + P_c H_{\rm eff} Q_c
\frac{1}{E^{(\rm h)}-Q_cH_{\rm eff}Q_c} Q_c H_{\rm eff} P_c
\label{heffhole}
\end{equation}
where $E^{(\rm h)}$ is the ground-state energy. Only those parts
of $H_{\rm eff}$ contribute to the last term to order $U^{-1}$
which generate a single doubly occupied site. Thus,
\begin{equation}
\tilde H^{(\rm h)} = P_c H_{\rm eff} P_c
- \frac{1}{U} P_c (H_t + H^{(1)}) Q_c (H_t + H^{(1)}) P_c
\label{hamtilde}
\end{equation}
(see Appendix~\ref{sec:general}). When terms of order $V^4$, as
well as terms involving three lattice sites are neglected the
second term in Eq.~(\ref{hamtilde}) becomes
\begin{equation}
\begin{array}{r c l}
\lefteqn{-\displaystyle
\frac{1}{U} P_c H_{\rm eff} Q_c H_{\rm eff} P_c=
\phantom{\sum_{jx}}} \hspace{7mm}
\\
&& \displaystyle\frac{2t^2}{U} \sum_{jx}^{\phantom{0}}
\left( {\bf s}_j {\bf s}_{j+x} - \frac{n_j n_{j+x}}{4} \right)
\\
&& + \displaystyle
\frac{t^2V^2(1-2r-2r^2)}{\epsilon_f^2 U(1+r)^2}
\sum_{i(0)}^{\phantom{0}}
\left( {\bf s}_0 {\bf s}_i -\frac{n_0 n_i}{4}\right)
\\
&& + \displaystyle
\frac{t^2V^2}{\epsilon_f^2 U (1+r)^2}
\sum_{i(0)}^{\phantom{0}}
{\bf S}_f( {\bf s}_0 n_i - {\bf s}_i n_0),
\end{array}
\label{virocc}
\end{equation}
where ${\bf s}_j$ is the spin at site $j$ and the summation
$i(0)$ is restricted over the nearest neighbors of site $0$,
only.

One notices that the first term on the right-hand side when
added to $P_c H_c P_c$ yields the well-known $t$-$J$ model,
i.e.,
\begin{equation}
\displaystyle H_{tJ} =
-t \sum_{jx,\sigma}
\hat c^\dagger_{j\sigma} \hat c^{\phantom{\dagger}}_{j+x \sigma}
+ \frac{2t^2}{U} \sum_{jx}
\left( {\bf s}_j {\bf s}_{j+x} - \frac{n_j n_{j+x}}{4} \right)
\end{equation}
with $\hat c^\dagger_{i\sigma} = c^\dagger_{i\sigma}
(1-n_{i\bar\sigma})$.  This is expected since the Hubbard
Hamiltonian $H_c$ reduces to $H_{tJ}$ in the strong correlation
limit. All terms depending on $V$ describe the interaction of
the magnetic impurity with the strongly correlated electrons. We
regroup the different terms in $\tilde H^{(\rm h)}$ and find
\begin{equation}
\tilde H^{(\rm h)} = \Delta\! E + H_{\rm kin}
+ H_{\rm H} + H_{\rm p} + H_{\rm K} + H' .
\label{hamtilde1}
\end{equation}
The different contributions are as follows. The term $\Delta\!
E$ describes a constant energy shift:
\begin{equation}
\displaystyle \Delta\! E
= \epsilon_f - \frac{\gamma V^2}{2(U-\epsilon_f)} ,
\end{equation}
where $\gamma= 1+Zt^2/(U-\epsilon_f)^2$ is again a
renormalization factor. The kinetic energy of the conduction
electrons is given by $H_{\rm kin}$, i.e.,
\begin{equation}
\begin{array}{r c l}
H_{\rm kin} & = & \displaystyle \sum_{jx,\sigma} T_j \left(
\hat c^\dagger_{j\sigma} \hat c^{\phantom{\dagger}}_{j+x \sigma}
+\hat c^\dagger_{j+x\sigma} \hat c^{\phantom{\dagger}}_{j\sigma}
\right)
\\
T_j & = & \displaystyle - \frac{t}{2} \left(
1 - \frac{V^2(2+2r+r^2)}{2\epsilon_f^2(1+r)^2} \delta_{j0}
\right) .
\end{array}
\end{equation}
One notices that due to the impurity the electron hopping is
reduced between site 0 and its nearest neighbors. The
interactions between the strongly correlated conduction
electrons are described by $H_{\rm H}$, which is of Heisenberg
type
\begin{equation}
\begin{array}{r c l}
H_{\rm H} & = & \displaystyle \sum_{jx} J_j
\left( {\bf s}_j {\bf s}_{j+x} - \frac{n_j n_{j+x}}{4}\right)
\\
J_j & = & \displaystyle \frac{2t^2}{U}  \left(
1 - \frac{V^2}{|\epsilon_f| (U-\epsilon_f)} \delta_{j0}
\right) .
\end{array}
\label{Heisenberg}
\end{equation}
Again, the interactions between site 0 and its nearest neighbors
are reduced due to the impurity. $H_{\rm p}$ is a one-particle
potential, which describes the attraction of a hole by the
impurity as well as its repulsion from a nearest neighbor site
\begin{equation}
\begin{array}{r c l}
H_{\rm p} & = & \displaystyle
-\frac{\eta V^2}{2|\epsilon_f|(1+r)} (1-n_0)
\phantom{\sum_{i(0)}}
\\
&& + \displaystyle
\frac{V^2t^2}{2\epsilon_f^2 U (1+r)^2} \sum_{i(0)} (1-n_i)
\\
\eta & = & \displaystyle
2 + r + \frac{Zt^2}{\epsilon_f^2 (1+r)^2}
\left( 2 + 7r + 7r^2 + r^3 \right) .
\end{array}
\end{equation}
The attraction of a hole by the impurity is intuitively
clear. If a hole is located at site 0 a virtual hop of the $f$
electron onto that lattice site creates a singly instead of a
doubly occupied state. The kinetic energy of the $f$ electron is
therefore increased. The last term shows that holes are weakly
repelled from the neighboring sites $i$.

The term $H_{\rm K}$ is the analogue of the spin-spin
interaction in the Kondo Hamiltonian
\begin{equation}
H_{\rm K} = \displaystyle
\frac{2\gamma V^2}{U-\epsilon_f} {\bf S}_f {\bf s}_0
-\frac{tV^2(2+r)}{U\epsilon_f(1+r)^2} {\bf S}_f
\sum_{i(0)} (\hat {\bf s}_{i0} + \hat {\bf s}_{0i}) ,
\label{kondoh}
\end{equation}
with $\hat {\bf s}_{0i} = 1/2 \sum_{\alpha\beta} \hat
c^\dagger_{i \alpha}
{\bbox{\sigma}}^{\phantom{\dagger}}_{\alpha\beta} \hat
c^{\phantom{\dagger}}_{0 \beta}$. Since site 0 is doubly
occupied in the virtual state, the dominant prefactor of the
first term is $2V^2/(U-\epsilon_f)$ rather than
$-2V^2/\epsilon_f$ as in the case of uncorrelated electrons
[cf.\ Eq.~(\ref{Uzero})]. The other terms arise from the
application of $L_t$ and indicate a delocalization of the
interaction.

Finally, there is a contribution $H'$ to $\tilde H^{(\rm h)}$,
which describes an interaction specific to the strong
correlation case
\begin{equation}
H' = \displaystyle\frac{2V^2t^2}{U \epsilon_f^2(1+r)^2}
\sum_{i(0)} {\bf S}_f
\left[ {\bf s}_i (1-n_0) - {\bf s}_0 (1-n_i) \right] .
\label{newinter}
\end{equation}
The first term implies an antiferromagnetic interaction between
the impurity and the nearest neighbor sites of site $0$ provided
there is a hole at site $0$. Note that in the half-filled case
the spins of the impurity and those of the nearest neighbor
sites $i$ are {\em ferromagnetically} aligned. The second term
can be considered as a correction to the Kondo-type
interaction. When an electron at site $0$ with spin ${\bf s}_0$
hops to an empty site $i$, the $f$ electron can hop virtually
onto site $0$ without creating a double occupancy.  Hence, the
antiferromagnetic spin exchange is decreased.

\subsection{Half-filled case}
\label{subsec:half}

In the case of half filling $n_j = 1$ for all $j$, and the
effective Hamiltonian~(\ref{hamtilde1}) reduces to
\begin{equation}
\begin{array}{r c l}
H_{\rm eff} & = & \Delta\!E + H_{\rm H} + H_{\rm K}
\phantom{\displaystyle\frac{Zt^2N_{\rm S}}{2U}\sum_{jx}}
\\
& = & \displaystyle
-\frac{Zt^2N_{\rm S}}{2U} + \epsilon_f
- \frac{\gamma V^2}{2(U-\epsilon_f)}
- \frac{ZV^2t^2}{2\epsilon_f U^2(1+r)}
\\
& & +\displaystyle
\sum_{jx} J_j {\bf s}_j {\bf s}_{j+x}
+ \frac{2\gamma V^2}{U-\epsilon_f} {\bf S}_f {\bf s}_0 ,
\end{array}
\label{halffilled}
\end{equation}
where $J_j$ is as in Eq.~(\ref{Heisenberg}). $N_{\rm S}$ denotes
the number of the lattice sites in the substrate. The conduction
electrons are described by an Heisenberg antiferromagnet with a
reduced Heisenberg exchange in the neighborhood of the
impurity. The $f$ spin couples via an antiferromagnet spin
exchange to the spin at the lattice site which is located
nearest to the $f$ orbital. If the (weak) $j$ dependence of
$J_j$ is neglected, the Hamiltonian (\ref{halffilled}) reduces
to that of Ref.~\onlinecite{NagaosaHI}.

\subsection{Electron doping}
\label{subsec:electron}

Next, we treat the case of a system with more than half filling
(electron doping). We proceed in close analogy to
Sec.~\ref{subsec:hole}, where the case of hole doping was
discussed. In the limit of large $U$, empty sites are excluded
here. Let $\tilde P_c$ denote the projector onto the
configuration space without empty sites and $\tilde Q_c =
1-\tilde P_c$. The effective Hamiltonian reduced to the space
without empty sites reads [cf.\ Eq.~(\ref{heffhole})]
\begin{equation}
\tilde H^{(\rm e)} = \tilde P_c H_{\rm eff} \tilde P_c
+ \tilde P_c H_{\rm eff} \tilde Q_c
\frac{1}{E^{(\rm e)}-\tilde Q_cH_{\rm eff} \tilde Q_c}
\tilde Q_c H_{\rm eff}\tilde P_c,
\end{equation}
where $E^{(\rm e)}$ is the ground-state energy. Only those parts
of $H_{\rm eff}$ contribute to the last term to order $U^{-1}$
which generate an empty site. Thus,
\begin{equation}
\tilde H^{(\rm e)}
= \tilde P_c H_{\rm eff} \tilde P_c
- \displaystyle \frac{1}{U} \tilde P_c
(H_t+H^{(1)}) \tilde Q_c (H_t+H^{(1)}) \tilde P_c .
\label{Hamelec}
\end{equation}
Again, we neglect terms of order $V^4$, as well as terms
involving three lattice sites. Using the relation $\tilde P_c
n_j \tilde P_c = 1 + n_{j\uparrow} n_{j\downarrow}$ and
regrouping the different terms in $\tilde H^{(\rm e)}$, we find
\begin{equation}
\tilde H^{(\rm e)} = \Delta\! E^{(\rm e)}
+ H_{\rm kin}^{(\rm e)} + H_{\rm H}^{(\rm e)}
+ H_{\rm p}^{(\rm e)} + H_{\rm K}^{(\rm e)} ,
\label{eldop}
\end{equation}
where
\begin{equation}
\begin{array}{r c l}
\Delta\! E^{(\rm e)} & = & \displaystyle \epsilon_f
- \frac{\gamma V^2}{2(U-\epsilon_f)} + U(N_{\rm c} - N_{\rm S})
\phantom{\sum_0^0}
\\
H_{\rm kin}^{(\rm e)} & = & \displaystyle
\sum_{jx,\sigma}^{\phantom{0}} T^{(\rm e)}_j
\left( \bar c^\dagger_{j\sigma}
\bar c^{\phantom{\dagger}}_{j+x\sigma}
+ \bar c^\dagger_{j+x\sigma}
\bar c^{\phantom{\dagger}}_{j\sigma} \right)
\\
H_{\rm H}^{(\rm e)} & = & \displaystyle
\sum_{jx}^{\phantom{0}}  J_j {\bf s}_j {\bf s}_{j+x}
\\
&& - \displaystyle\sum_{jx}^{\phantom{0}} \frac{J_j}{4}
(1-n_{j\uparrow}n_{j\downarrow})
(1-n_{j+x\uparrow}n_{j+x\downarrow})
\\
H_{\rm p}^{(\rm e)} & = & \displaystyle
\frac{\gamma V^2}{2(U-\epsilon_f)} n_{0\uparrow} n_{0\downarrow}
\phantom{\sum_0^0}
\\
H_{\rm K}^{(\rm e)} & = & \displaystyle
\frac{2\gamma V^2}{U-\epsilon_f} {\bf S}_f {\bf s}_0
+ \frac{tV^2}{(U-\epsilon_f)^2} {\bf S}_f
\sum_{i(0)}^{\phantom{0}}
(\bar {\bf s}_{0i} + \bar {\bf s}_{i0}).
\end{array}
\end{equation}
$N_{\rm S}$ is the number of lattice sites in the substrate,
$N_{\rm c}$ is the number of conduction electrons, so that there
are $N_{\rm c} - N_{\rm S}$ doubly occupied sites, which
contribute with an energy $U(N_{\rm c} - N_{\rm S})$ to the
energy shift $\Delta\! E^{(\rm e)}$.

As in the case of hole doping, the second and third term in
Eq.~(\ref{eldop}), $H^{(\rm e)}_{\rm kin} + H^{(\rm e)}_{\rm
H}$, define a $t$-$J$ model describing the conduction electrons
with parameters, which are modified in the neighborhood of the
impurity:
\begin{equation}
\begin{array}{r c l}
T^{(\rm e)}_j & = & -\displaystyle\frac{t}{2}
\left( 1 - \frac{V^2}{2(U-\epsilon_f)^2} \delta_{j0} \right)
\\
J_j & = & \displaystyle \frac{2t^2}{U} \left(
1 - \frac{V^2}{|\epsilon_f| (U-\epsilon_f)} \delta_{j0}
\right) .
\end{array}
\end{equation}
In the case of electron doping, double occupancies hop rather
than holes: $\bar c^\dagger_{j\sigma} = c^\dagger_{j\sigma}
n_{j\bar\sigma}$. Since the $f$ electron can gain kinetic energy
only if site $0$ is singly occupied, double occupancies are
repelled from site $0$. This is described by the ``potential'',
$H_{\rm p}^{(\rm e)}$. The Kondo-type interaction, $H_{\rm
K}^{(\rm e)}$, between the $f$ spin and the spin of the
conduction electrons consists of two parts. The local part is
the same as in the case of hole doping, cf.\
Eq.~(\ref{kondoh}). The prefactor of the second term is modified
and we introduced $\bar {\bf s}_{0i} = 1/2 \sum_{\alpha\beta}
\bar c^\dagger_{i \alpha}
{\bbox{\sigma}}^{\phantom{\dagger}}_{\alpha\beta} \bar
c^{\phantom{\dagger}}_{0 \beta}$. However, in the case of
electron doping there are no interactions of the $f$ spin with
the conduction electrons corresponding to $H'$, cf.\
Eq.~(\ref{newinter}).

Finally, we mention that in the half-filled case
\hbox{$n_{j\uparrow} n_{j\downarrow} =0$}, and the
Hamiltonian~(\ref{eldop}) reduces to that of
Eq.~(\ref{halffilled}).

\section{Summary}
\label{sec:summary}

In this paper, we considered a magnetic impurity which interacts
weakly by hybridization with a system of strongly correlated
electrons. This contrasts the situation treated in the Anderson
impurity model where the impurity interacts with {\em free}
conduction electrons. We described the strong correlations among
the conduction electrons by a Hubbard Hamiltonian and performed
a canonical transformation to eliminate the charge degrees of
freedom at the impurity site. For vanishing Hubbard repulsion
$U$, this procedure was shown to be equivalent to the
conventional Schrieffer-Wolff transformation and yields the
Kondo Hamiltonian.

Of particular interest is the limit $U\gg t$. Here, we
additionally reduced the charge degrees of freedom of the
conduction electrons by a degenerate perturbation expansion
which, in the absence of the magnetic impurity, leads to the
$t$-$J$ model. The corrections we find are due to the
impurity. In the case of half filling, the conduction electrons
are described by a Heisenberg Hamiltonian with an exchange
coupling constant, which is slightly modified in the
neighborhood of the impurity. Furthermore, there is an exchange
interaction between the impurity spin and the spin of the
conduction electron at the lattice site closest to the impurity,
i.e., site $0$. The coupling constant depends on $U$ and is,
therefore, different than in the usual Kondo Hamiltonian.

When the system is doped new terms arise. Since we strictly
forbid double occupancies of the impurity site the electron-hole
symmetry is broken and we have to treat the cases of hole and
electron doping separately. If the system is doped with
electrons, doubly occupied sites are repelled by the impurity
since otherwise the exchange becomes impossible.

When we dope the system with holes, the conduction electrons
move like in a $t$-$J$ model. The holes are attracted towards
the impurity, which reflects the gain in kinetic energy if the
$f$ electron hops without creating a doubly occupied
site. Additionally, we find a new type of interaction: Provided
a hole is located at the lattice site $0$ there is an
antiferromagnetic exchange of the impurity spin with the
conduction electron spin at the nearest neighbor sites of
$0$. In the undoped system these spins are ferromagnetically
aligned.

\widetext

\appendix
\section{Effective Hamiltonian}
\label{sec:general}

In the following, we derive the effective
Hamiltonian~(\ref{heff2}) which is obtained from the starting
Hamiltonian~(\ref{ham}) after performing the canonical
transformation~(\ref{hcan}) that eliminates the charge degrees
of freedom of the $f$ orbital and projecting onto the space with
$n_f = 1$. We begin by listing the terms to lowest order in the
expansion~(\ref{expand}) of $1/(\epsilon_f-L_c)$. The
zeroth-order term ($\nu=0$) reads
\begin{equation}
\frac{1}{\epsilon_f - L_U} c^\dagger_{0\sigma} =
\frac{1}{\epsilon_f}
\left( 1-\frac{1}{1+r} n_{0\bar\sigma} \right)
c^\dagger_{0\sigma} .
\label{ord0}
\end{equation}
For $\nu=1$ one obtains
\begin{equation}
\begin{array}{r c l}
\displaystyle\frac{1}{\epsilon_f-L_U} L_t
\frac{1}{\epsilon_f-L_U} c^\dagger_{0\sigma} & = &
\displaystyle \frac{-t}{\epsilon_f^2} \sum_{i(0)}
\left[
\left( 1-\frac{1}{1+r} n_{i\bar\sigma} \right)
\left(1-\frac{1}{1+r} n_{0\bar\sigma} \right)
c^\dagger_{i\sigma}
\right.
\\
&& \displaystyle\left.~
- \frac{1}{1+r} \left(1-\frac{1}{1+r} n_{i\sigma}\right)
c^\dagger_{i\bar\sigma} c^{\phantom{\dagger}}_{0\bar\sigma}
c^\dagger_{0\sigma}
+ \frac{1}{(1+r)^2} \left( r + n_{i\sigma} \right)
c^\dagger_{0\bar\sigma} c^{\phantom{\dagger}}_{i\bar\sigma}
c^\dagger_{0\sigma} \right]
\end{array}
\label{ord1}
\end{equation}
and for $\nu=2$
\begin{equation}
\displaystyle \frac{t^2}{\epsilon_f^3} \sum_{i(0)} \left[
\left( 1 - \frac{(1+2r)^2}{(1+r)^3} n_{0\bar\sigma}
+\frac{r}{(1+r)^2} n_{i\bar\sigma} \right) c^\dagger_{0\sigma}
\displaystyle  - \frac{r}{(1+r)^2}
\left( c^\dagger_{i\bar\sigma}
c^{\phantom{\dagger}}_{0\bar\sigma} c^\dagger_{i\sigma}
- c^\dagger_{0\bar\sigma} c^{\phantom{\dagger}}_{i\bar\sigma}
c^\dagger_{i\sigma} \right) \right] .
\label{ord2}
\end{equation}
The summations $i(0)$ are restricted over nearest neighbor sites
$i$ of site 0 and in Eq.~(\ref{ord2}) we have neglected
contributions from next-nearest neighboring sites. When the
Eqs.~(\ref{ord0}--\ref{ord2}) are set into Eq.~(\ref{heff1}) one
obtains $H_{\rm eff}$ in the form
\begin{equation}
H_{\rm eff}= H_c + \epsilon_f + H^{(0)} + H^{(1)} + H^{(2)} .
\end{equation}
The different contributions are
\begin{equation}
H^{(0)} = \displaystyle \frac{V^2}{\epsilon_f} -
\frac{V^2(2+r)}{2\epsilon_f(1+r)} n_0
+ \frac{V^2}{\epsilon_f(1+r)} n_{0\uparrow} n_{0\downarrow}
+ \frac{2V^2}{U(1+r)} {\bf S}_f {\bf s}_0 ,
\end{equation}
\begin{equation}
\begin{array}{r c l}
H^{(1)} & = & \displaystyle
\frac{tV^2}{4\epsilon_f^2(1+r)^2} \sum_{i(0)\sigma}
\left(c^\dagger_{i\sigma} c^{\phantom{\dagger}}_{0\sigma}
+ c^\dagger_{0\sigma} c^{\phantom{\dagger}}_{i\sigma} \right)
\left( 2 + 2r + r^2 - (2+r) n_{0\bar\sigma} - (3+r)
n_{i\bar\sigma} + 3n_{0\bar\sigma} n_{i\bar\sigma}
\right)
\\
&& + \displaystyle \frac{tV^2}{2\epsilon_f^2(1+r)^2}
S_f^z \sum_{i(0)\sigma} (-)^\sigma \left(
c^\dagger_{i\sigma} c^{\phantom{\dagger}}_{0\sigma}
+ c^\dagger_{0\sigma}c^{\phantom{\dagger}}_{i\sigma}
\right)
\left( r(r+2) + (1-r) n_{i\bar\sigma}
- r n_{0\bar\sigma} - n_{0\bar\sigma} n_{i\bar\sigma}\right)
\\
&& + \displaystyle\frac{tV^2}{2\epsilon_f^2(1+r)^2}
\sum_{i(0)} \left( S_f^+ s_{i0}^- + S_f^- s_{0i}^+ \right)
\left( r(r+2) + (1-r) n_{i\uparrow}
- r n_{0\downarrow} -  n_{0\downarrow} n_{i\uparrow} \right)
\\
&& + \displaystyle\frac{tV^2}{2\epsilon_f^2(1+r)^2}
\sum_{i(0)} \left( S_f^+ s_{0i}^- + S_f^- s_{i0}^+ \right)
\left( r(r+2) + (1-r) n_{i\downarrow}
- r n_{0\uparrow} - n_{0\uparrow} n_{i\downarrow} \right) ,
\end{array}
\end{equation}
and
\begin{equation}
\begin{array}{r c l}
H^{(2)} & = & \displaystyle \frac{ZV^2t^2}{\epsilon_f^3}
-\frac{V^2t^2}{2U\epsilon_f^2(1+r)^2} \sum_{i(0)}
\left( c^\dagger_{i\uparrow} c^\dagger_{i\downarrow}
c^{\phantom{\dagger}}_{0\downarrow}
c^{\phantom{\dagger}}_{0\uparrow}
+  c^\dagger_{0\uparrow} c^\dagger_{0\downarrow}
c^{\phantom{\dagger}}_{i\downarrow}
c^{\phantom{\dagger}}_{i\uparrow} \right)
\\
&& - \displaystyle
\frac{ZV^2t^2 \left( 2 + 7r + 7r^2 + r^3 \right)}
{2\epsilon_f^3(1+r)^3} n_0
-\frac{V^2t^2}{2U\epsilon_f^2(1+r)^2} \sum_{i(0)} n_i
+\frac{ZV^2t^2 (1+2r)^2}{\epsilon_f^3(1+r)^3}
n_{0\uparrow} n_{0\downarrow}
\\
&& - \displaystyle\frac{V^2t^2}{U\epsilon_f^2(1+r)^2}
\sum_{i(0)} \left( {\bf s}_0 {\bf s}_i - \frac{n_0 n_i}{4}
\right)
\\
&& - \displaystyle\frac{V^2t^2}{U\epsilon_f^2(1+r)^2}
\sum_{i(0)} {\bf S}_f {\bf s}_0
\left( \frac{2(1+r-r^2)}{1+r} - n_i \right)
+ \frac{V^2t^2}{U\epsilon_f^2(1+r)^2}
\sum_{i(0)} {\bf S}_f {\bf s}_i \left( 2-n_0 \right) .
\end{array}
\end{equation}
The Hamiltonian $H_{\rm eff}$ is discussed in several limiting
cases in Secs.~\ref{sec:smallU} and \ref{sec:largeU}.

\narrowtext

\end{document}